\documentstyle[12pt,psfig]{article}

\def\ni{\noindent}
\def\be{\begin{equation}}
\def\ee{\end{equation}}

\begin{document}
\setlength{\textheight}{7.7truein}  

\thispagestyle{empty}
\centerline{\bf Simple Simulational Model for}
\vspace*{0.035truein}
\centerline{\bf Stocks Markets}
\vspace*{0.37truein}
\centerline{\footnotesize JUAN R. SANCHEZ
\footnote{Email: jsanchez@fi.mdp.edu.ar}}
\baselineskip=12pt
\centerline{\footnotesize\it Departamento de F\'{\i}sica,
Facultad de Ingenier\'{\i}a}
\baselineskip=10pt
\centerline{\footnotesize\it Universidad Nacional de Mar del Plata}
\baselineskip=10pt
\centerline{\footnotesize\it Av. J.B. Justo 4302, 7600 Mar del Plata,
Argentina}
\vspace*{0.225truein}

\vspace*{0.25truein}
\abstract{A new model for stocks markets using integer values for
each stock price is presented. In contrast with previously reported 
models, the variables used in the model are not of binary type, 
but of more general integer type.
It is shown how the behavior of the noisy and fundamentalists traders
can be taken into account simultaneously in the time evolution 
of each stock price. The simulated time series 
is analysed in different ways order to compare parameters with those of real 
markets.}{}{}

\vspace*{12pt}     

\vspace*{1pt}
Recently, several computational models trying to represent the behavior of
actual stocks markets have been presented.~\cite{1,2,3}
>From these results it seems to be a well established fact that, in order to 
obtain a good representation of the time evolution of actual markets, 
two type of traders agents must be included.
On one side there are the so called {\it noisy} traders which are supposed to
follow the {\it local} (in space and time) trend of the market. 
The noisy traders, also called followers, place their buy or sell orders 
on a given stock following
the behavior of other (related) stocks. This kind of attitude seems 
to be an almost evident practice for someone trying to operate
within a market.
But, on the other hand, there are also {\it fundamentalists} traders which are 
considered to be responsible of the market turn offs. 
These kind of traders are supposed to know something more about the
market and then are able to develop some kind of more sophisticated 
strategy to operate.
They can take actions to buy or sell according to other indicators;
the markets {\it fundamentals}. These indicators could depend on other type
of information which usually may come from {\it outside} of the system.
In previous models the two types of agents are represented in a variety
of ways.~\cite{2,3} 
For instance, in reference 2 the influence of the fundamentalists is 
modeled by including a term that takes into account the
{\it tendency} in the price changes. This effect stabilizes the prices.
On the other hand, in reference 3 the two type of traders are included 
as separate entities.
Then, it is realistic to think that both type of
behaviors are acting at the same time and influence the way in which 
a specific stock price change. 
>From the point of view of simulations, another common characteristic of many
of the models just presented seems to be the use of Ising like variables 
that represent the buy or sell attitude of market agents.

Taking into account all the above mentioned characteristics, 
a somewhat different modeling approach is presented  here.
In principle, the model is based in previously reported
models of opinion evolution in a closed community.~\cite{1} 
But, instead of using Ising spins, here the community is modeled by 
a vector of stock prices $\mathbf{x}$ having $N$ integer 
valued ``Potts'' components $x_i$, each one 
representing the price of a market asset (in arbitrary units). 
Then, associated with each $x_i$ is a {\it direction of movement} value
$v_i$. The components $v_i$ form a vector $\mathbf{v}$. 
This components are of Ising type, i.e., 
they can take two values $v_i=+1$ and $v_i=-1$.
The vectors $\mathbf{x}$ and $\mathbf{v}$
evolve in time according to the following dynamical rules.
A randomly chosen component $x_i$ is updated according to the equation
\begin{equation}\label{eq1}
x_i(t+\Delta t)=x_i(t)+v_k(t) \:;
\end{equation}
\ni while for the corresponding $v_i$ components the following 
equation is valid
\begin{equation}\label{eq2}
v_i(t+\Delta t) = \left\{
\begin{array}{rl}
v_k(t) & {\mathrm {if}} \:\: |x_i(t)| < X_{th} \\
 & \\
-v_i(t) &  {\mathrm {if}} \:\: |x_i(t)| > X_{th} \: .
\end{array} \right .
\end{equation} 
\ni The value of $v_k$ is obtained by choosing at random among 
the direction values of each one of the
{\it neighbors}, i.e., $v_k = v_{i-1}$ or $v_k = v_{i+1}$ with equal
probability.
In principle, the algorithm described by the equations~\ref{eq1} 
and \ref{eq2} takes into account
the influence of the noisy traders which follow the trend of 
related stocks in order to buy ($v_i = +1$) or to sell 
($v_i = -1$) an specific asset. 
However, it is not reasonable to think that the prices $\mathbf{x}$ 
could take arbitrary positive or negative values. There is no actual market
following a given trend for ever. Then in order to
take into account the influence of the fundamentalists traders, a
threshold $X_{th}$ is established for the {\it absolute} value of 
each $x_i$. As it can be seen in equation~\ref{eq2}, if at any time 
$|x_i| > X_{th}$ the corresponding direction of movement $v_i$ is 
{\it reversed}, $v_i \rightarrow -v_i$. 
This reversal procedure simulates the influence of the
fundamentalists traders which, when the absolute value of a stock 
reaches a value $X_{th}$, consider that the price is low enough so is time 
to buy or it is high enough and then is time to sell. 
In equations~\ref{eq1} and \ref{eq2}, $\Delta t$ is proportional 
to $N^{-1}$ so the simulation time 
is incremented by one when all the stocks have, on average, 
the chance to evolve once.

In order to analyze the behavior of the model, two representative indexes 
of the market, the mean value time series
\begin{equation}\label{eq3}
x_M(t) = \frac{1}{N} \: \sum_{i=1}^{N} x_i(t)
\end{equation}
and the corresponding {\it returns} or change of price, defined here as
\begin{equation}\label{eq4}
r(t) = x_M(t) - x_M(t-1)
\end{equation} 
are investigated by Monte Carlo simulations.
$N=1024$ and $X_{th}=30$ where used as typical parameters
and $T \cong 20000$ Monte Carlo steps (MCS) where used in order to obtain 
most of the reported results. 
To avoid the existence of initial correlations the simulations 
are started with the components of the vector $\mathbf{x}$ distributed 
randomly
between $-10$ and $10$ and with the vector $\mathbf{v}$ in a complete 
antiferromagnetic state.
 
A typical path of the simulated price series $x_M(t)$ 
is shown in Fig. 1. The sequence shows the well known noisy shape.
The market turn offs have been indicated in the figure. This turn offs
have a self-organized character, since they result from the 
time evolution of the model and their occurrence cannot be explicitly 
predicted from the dynamic equations.
Previous analysis of actual markets time series suggest that the 
probability distribution of daily returns are not of Gaussian type,
but they are {\it fat-tailed}.~\cite{4} This characteristic can 
be noticed if the probability distribution function (PDF) of the returns
is plotted on the scale of the cumulative Gaussian distribution 
function. Normal distributions appear as straight lines in such 
representation while the fat-tail of other type of distributions result in
a departure of the straight line. For the model presented here,
the PDF of the price returns is
plotted in Fig. 2 and the tails are clearly visible .

According to previous results, it is known that actual markets time series
cannot be modeled by series of independent and identically distributed
realizations of random variables with a given distribution (fat-tailed
or not). This is a consequence of the existence of non-stationarity in
the process that generate the series. It has been shown 
that a direct method exists in order to detect the non-stationarity, it is 
the calculation of the autocorrelation function~\cite{3,5} 
\begin{equation}\label{eq5}
{\mathbf{acf}}(r,t') = 
\frac{\sum_{t=t'+1}^{T}(r_t - \overline{r})
(r_{t-t'} - \overline{r})}{\sum_{t=t'+1}^{T}(r_t - \overline{r})^2} \:\:,
\end{equation}
of the {\it absolute} value of the returns $|r(t)|$. For Brownian motion
the ${\mathbf{acf}}(r,t')$ of $r(t)$, $r^2(t)$ and $|r(t)|$ fluctuates
around zero for $t' > 1$. In Fig. 3 the ${\mathbf{acf}}(r,t')$ for 
the $r(t)$ time series generated by the model is plotted up to $t'=50$. 
The horizontal lines represent the $0.95$ confidence interval of a 
Brownian random walk.
Clearly, it is shown that the simulated series has a certain degree of 
non-stationarity very similar of those just reported for actual markets 
time series.

Another parameter that can reflect the deviation of a time series from
the Gaussian distribution is the excess kurtosis, defined as
\begin{equation}\label{eq6}
\kappa = \frac{\mu_4}{\sigma^4} - 3
\end{equation}
\ni where $\mu_4$ is the fourth central moment and $\sigma$ is the standard
deviation of the series under study. 
$\kappa$ is defined to be zero for a normal distribution,
but ranges between $2$ and $50$ for daily returns of actual stock markets
data have been reported.~\cite{6,7}
An average value of $\overline \kappa = 2.69$
was obtained when the kurtosis is
calculated on the absolute values of the returns. 
The value $\overline \kappa$ was obtained by averaging over
$100$ independent runs of $T=10000$ MCS.

Finally, the simulated series was analyzed using the Hurst R/S 
method.~\cite{4} 
The R/S analyses begins with dividing the time series in segments of
equal length and normalizing the data in each segment by subtracting 
the sample mean. Then, the rescaled range (range/standard deviation) 
is log-log plotted against the segment size. By linear regression of the 
plot the Hurst exponent $H$ is obtained. The values of the exponent
reflect some characteristics of the series: for $H > 0.5$ the series is said
to be persistent, if $H=0.5$ the series represent a normal distributed 
random walk, while for $H < 0.5$ the series is considered to have anti-persistent
characteristics. 
The R/S analysis for the
price and returns series is presented in Fig. 4. 
The straight lines through the
points indicates that the Hurst exponent is $H_x = 0.885 \pm 0.01$ for the
$x_M(t)$ series and $H_r = 0.569 \pm 0.01$ for the $r(t)$ series. Since
the values are greater than $0.5$ both series show a
persistent character which results
from the existence of long term correlations in both
distributions. In particular, the value of $H_r$ is very close to the
value of the Hurst exponent calculated on the returns of the
USD/DEM exchange rate.

Although several other analysis could be made, the above presented results 
show that  the simulated time series obtained from the operation of the dynamic 
rules~\ref{eq1} and \ref{eq2} can be considered
as a good approximation of the behavior of actual stock markets. 
Also, the discrete nature of the price variables follow closely the same
characteristic of actual stocks prices. Further refinement could be made on the
model in order to reproduce more closely some particular characteristic
of a given market.

This work was partially supported by a research grant from Universidad
Nacional de Mar del Plata (Mar del Plata, Argentina).

\newpage

\newpage
\begin{figure}[htbp] 
\vspace*{13pt}
{\psfig{file=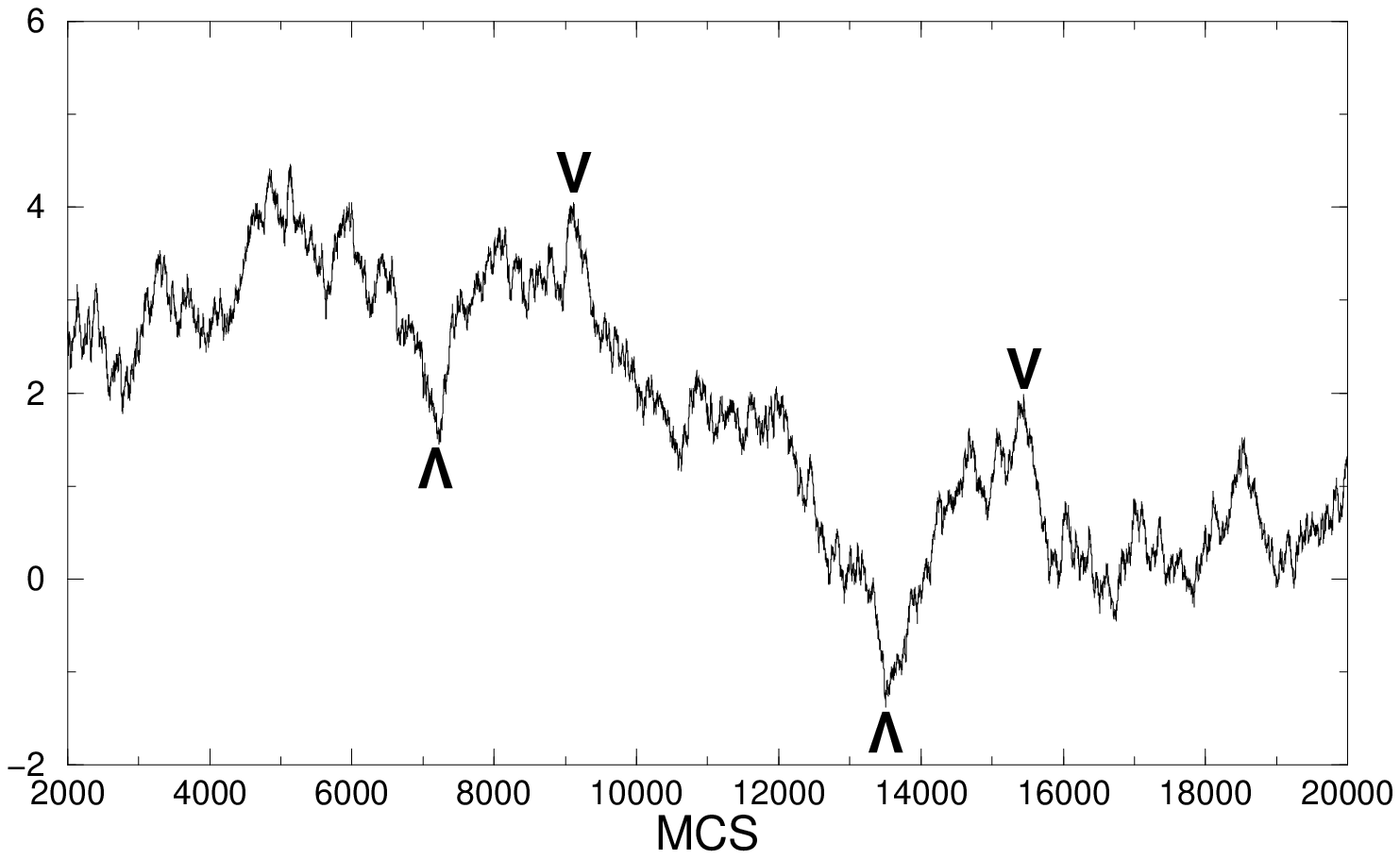}} 
\vspace*{13pt}
\caption{A typical path of the simulated price process $x_M(t)$.
The self-organized market turn offs are indicated by the {\bf V} symbols.}
\end{figure}

\begin{figure}[htbp] 
\vspace*{13pt}
{\psfig{file=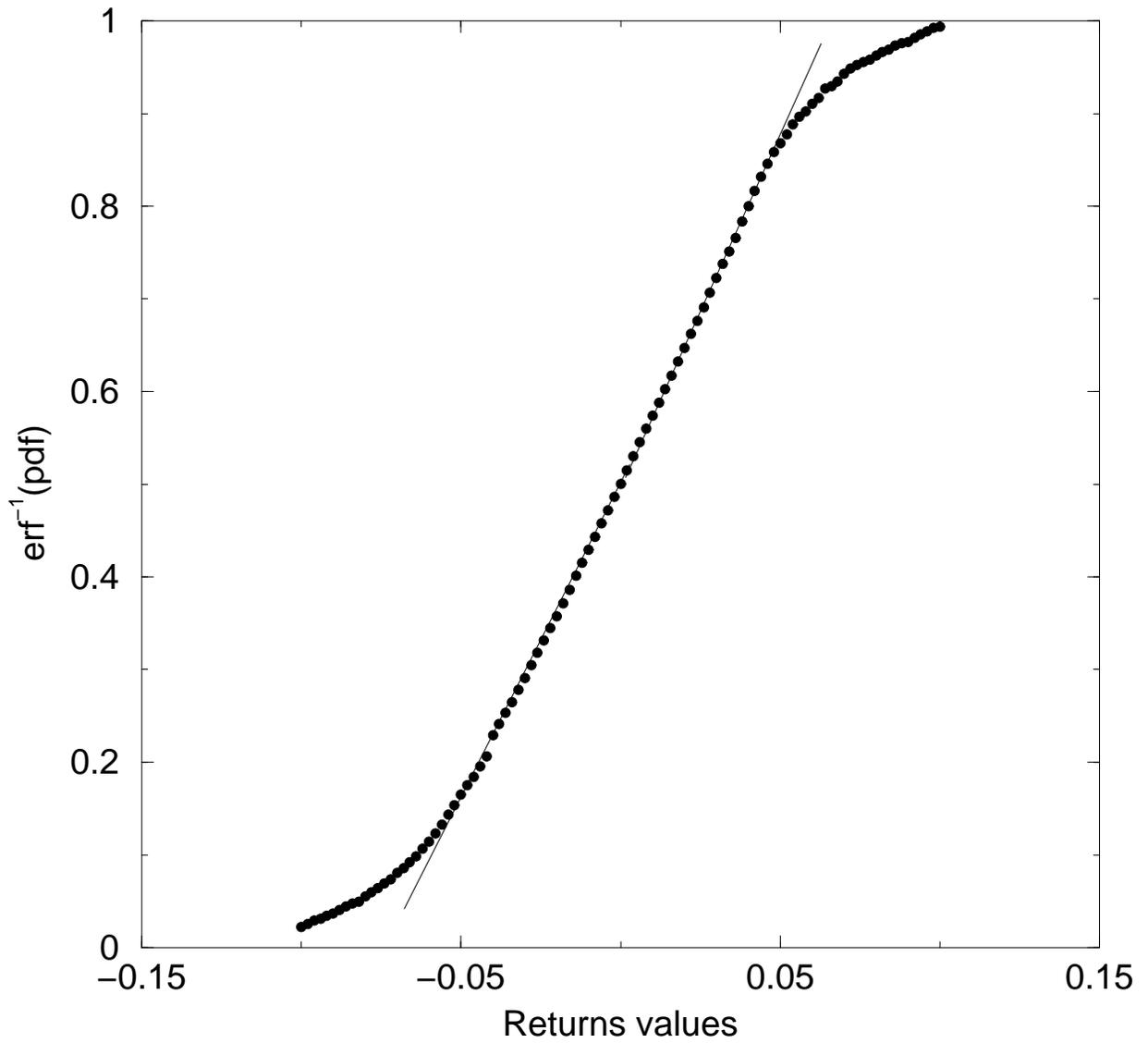}} 
\vspace*{13pt}
\caption{Probability distribution function of the price returns
plotted on an inverse error function scale.
The departure from the straight line show the existence of fat-tails
in the returns distribution.} 
\end{figure}

\newpage
\begin{figure}[htbp] 
\vspace*{13pt}
{\psfig{file=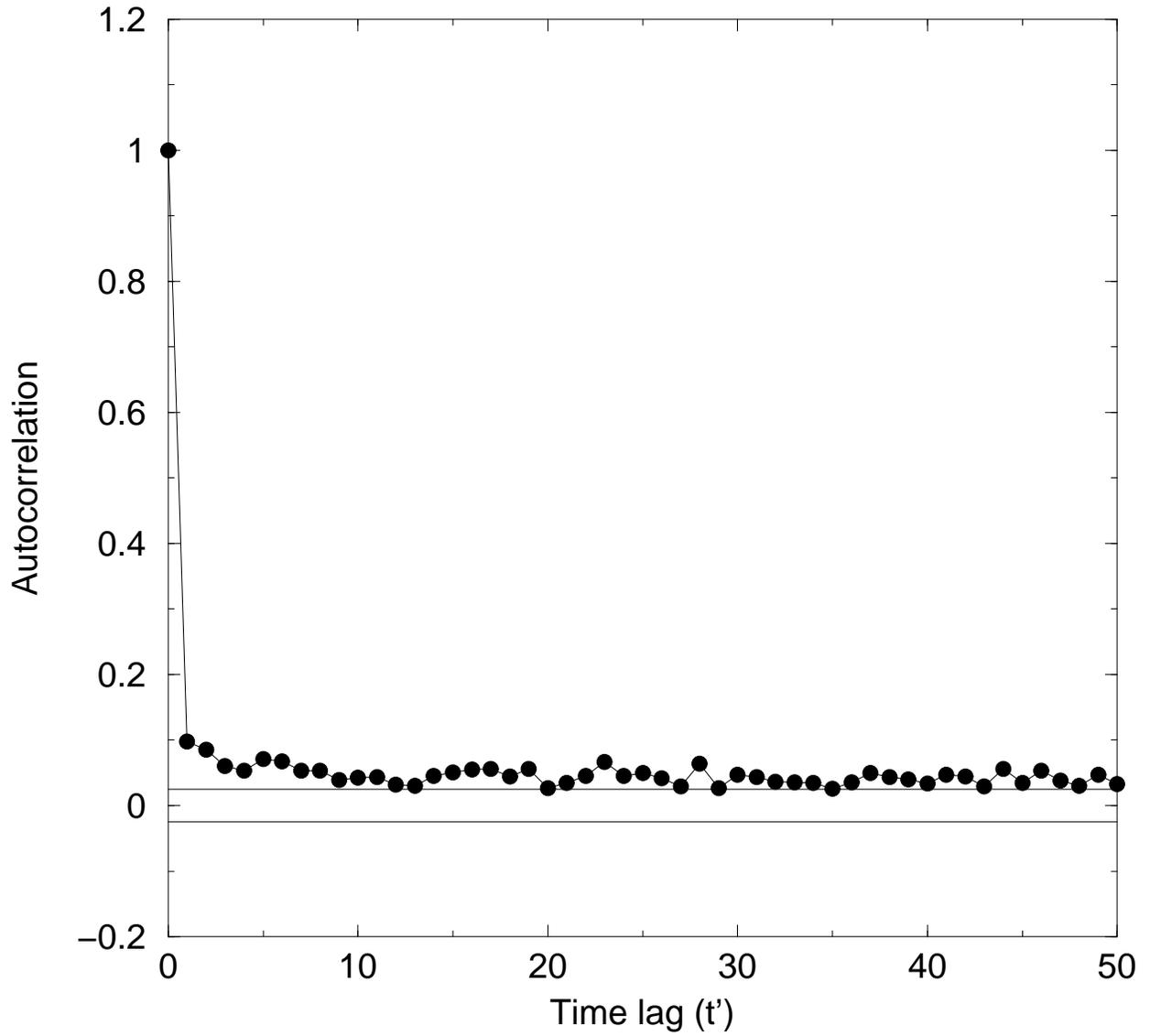}} 
\vspace*{26pt}
\caption{Autocorrelation function ${\mathbf{acf}}(r,t')$ 
(see equation~\ref{eq4}) for the absolute returns $|r(t)|$
generated by the model plotted up to $t'=50$. The horizontal
lines represent the $0.95$ confidence interval of a Brownian random walk.}
\end{figure}

\newpage
\begin{figure}[htbp] 
\vspace*{13pt}
{\psfig{file=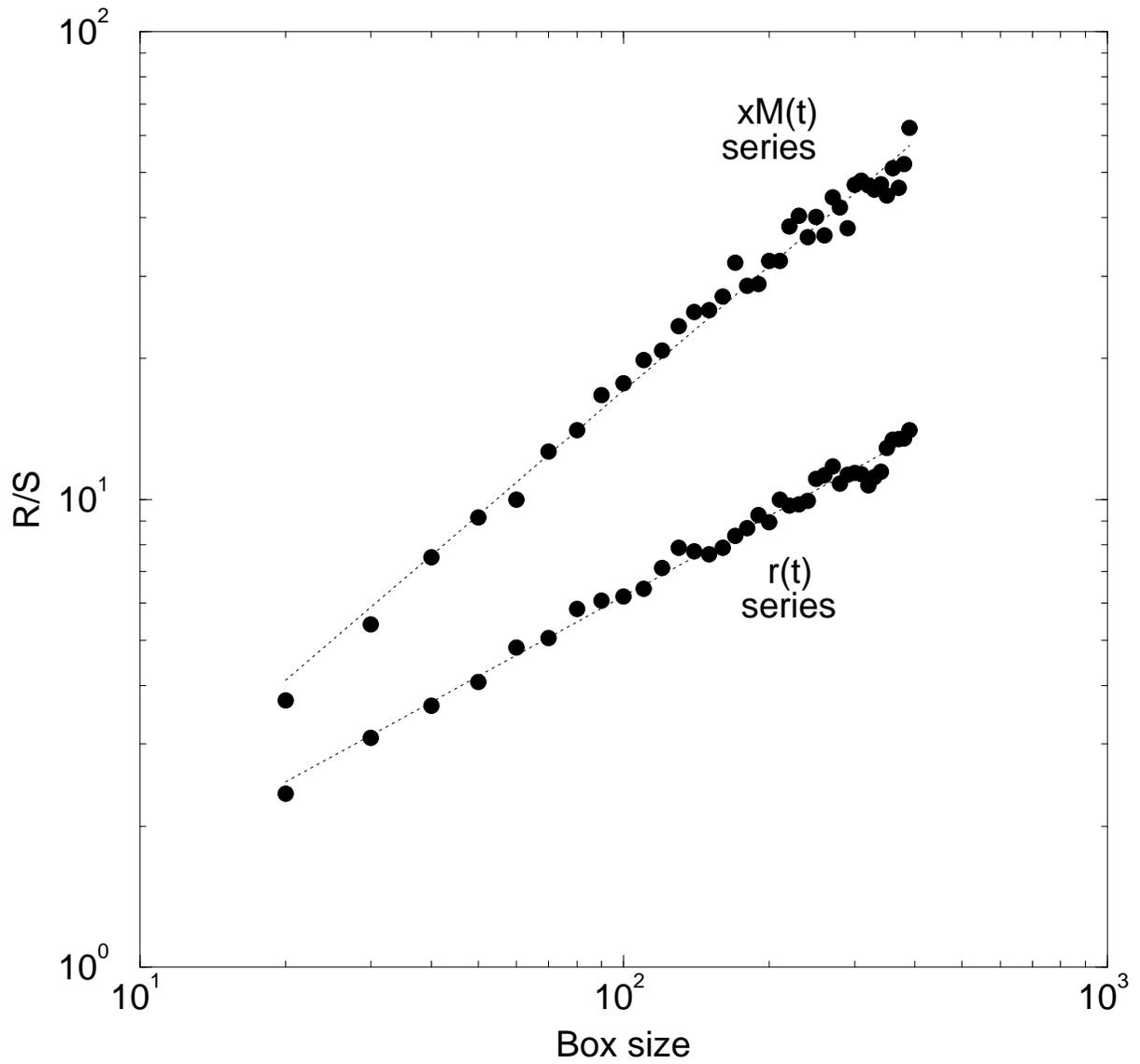}} 
\vspace*{13pt}
\caption{R/S analysis for the
price ($xM$) and returns ($r$) time series.}

\end{figure}

\end{document}